\documentclass[12pt]{article}

\usepackage{amsthm,amsfonts,amssymb,amsmath}
\usepackage{xspace}
 \usepackage{bm}
\usepackage{authblk}
\usepackage{graphicx}

\usepackage{natbib}

\newtheorem{remark}{Remark}

\newcommand{\A}{\ensuremath{\mathbb{A}}\xspace}
\newcommand{\B}{\ensuremath{\mathbb{B}}\xspace}
\newcommand{\R}{\mathbb{R}}
\newcommand{\N}{\mathbb{N}}
 \newcommand{\tsp}{\mathsf{s}}

 \newcommand{\ba}{\mathbf{a}}
 \newcommand{\bb}{\mathbf{b}}
 \newcommand{\bx}{\mathbf{x}}

 \newcommand{\bydef}{:=}
\newcommand{\rd}{\mathrm{d}}
\newcommand{\e}{\mathrm{e}}
\newcommand{\fp}{\mathrm{fp}}

\title{Population dynamics and games of variable size}

\date{\today}

\author[1]{Matheus Hansen\footnote{mh.francisco@fct.unl.pt}}
\author[1,2]{Fabio A. C. C. Chalub\footnote{Corresponding author: facc@fct.unl.pt}}

\affil[1]{{Center for Mathematics and Applications (NOVA Math), NOVA FCT, Universidade NOVA de Lisboa}, {Quinta da Torre}, {2829-516}, {Caparica}, {Portugal}}

\affil[2]{{Department of Mathematics, NOVA FCT, Universidade NOVA de Lisboa}, {Quinta da Torre}, {2829-516}, {Caparica}, {Portugal}}

\begin{document}

\maketitle

%
%
%
%
%
%
%
%
%

\begin{abstract}
This work introduces the concept of Variable Size Game Theory (VSGT), in which the number of players in a game is a strategic decision made by the players themselves. We start by discussing the main examples in game theory: dominance, coexistence, and coordination. We show that the same set of pay-offs can result in coordination-like or coexistence-like games depending on the strategic decision of each player type.  We also solve an inverse problem to find a $d$-player game that reproduces the same fixation pattern of the VSGT. In the sequel, we consider a game involving prosocial and antisocial players, i.e., individuals who tend to play with large groups and small groups, respectively. In this game, a certain task should be performed, that will benefit one of the participants at the expense of the other players. We show that individuals able to gather large groups to perform the task may prevail, even if this task is costly, providing a possible scenario for the evolution of eusociality. The next example shows that different strategies regarding game size may lead to spontaneous separation of different types, a possible scenario for speciation without physical separation (sympatric speciation). In the last example, we generalize to three types of populations from the previous analysis and study compartmental epidemic models: in particular, we recast the SIRS model into the VSGT framework: Susceptibles play 2-player games, while Infectious and Removed play a 1-player game. The SIRS epidemic model is then obtained as the replicator equation of the VSGT. We finish with possible applications of VSGT to be addressed in the future.
\end{abstract}

\newcommand{\sep}{$|$}
\textbf{keywords}: 
Game theory \sep Fixation \sep Evolution of eusociality \sep  Speciation \sep Epidemic models.


\maketitle

\section{Introduction}

Game theory is the part of mathematics that models strategic behavior~\citep{vonNeumann_Morgenstern}. A symmetric $d$-player game is defined by $d$ individuals (\emph{players}), that have to choose among $n$ \emph{strategies} $\mathbf{e}_{i}$, $i=1,\dots,n$, and as a result each one receive a certain \emph{pay-off}. A common assumption when modeling economic behavior is so-called \emph{rationality}: each player acts to maximize his or her pay-off function. A set of strategies in which no rational player has incentives to deviate from his or her strategy is called a \emph{Nash equilibrium}. It is possible to prove that, if players are allowed to choose their strategies according to certain probability distributions, there is always at least one Nash equilibrium in every game defined as above. More general formulations of games, in particular with applications to modeling economic behavior, can be found in~\citep{Gintis} and references therein.

In biology, it is customary to replace the rationality assumption with a given dynamics. In particular, a population of individuals of $n$ different types where each type adopts one of the strategies $\mathbf{e}_{i}$, $i=1,\dots,n$ is considered. The fraction of individuals of type $i$ at time $t\ge 0$ is given by $x_{i}(t)\in[0,1]$, and the \emph{state} of the population is given by $\mathbf{x}(t)=(x_1(t),\dots,x_n(t))\in  S^{n-1}:=\{\mathbf{z}\in\R^{n}_{+}, \sum_{i=1}^nz_i=1\}$, the $n-1$-dimensional \emph{simplex}, for all $t\ge0$. Pay-off functions, which are frequently identified as \emph{fitness} in the biological literature, are smooth functions $\psi_{i}:S^{i-1}\to\R$, $i=1,\dots,n$; more precisely, we say that, at state $\mathbf{x}$, $i$-type individuals have fitness $\psi_{i}(\mathbf{x})$. The state of the population evolves according to a certain given dynamics, from which the most popular is the \emph{replicator dynamics}, introduced in~\cite{TaylorJonker_1978}: $x_i'=x_i\left(\psi_{i}(\mathbf{x})-\overline{\psi}(\mathbf{x})\right)$, where $\overline{\psi}(\bx):=\sum_{i=1}^nx_i\psi_{i}(\bx)$ is the average pay-off; cf.~\citep{HofbauerSigmund} and references therein.

One common limitation of the previous description is that the number of players $d$ in a given game is assumed fixed. However, the number of individuals we interact with at any given period of time in our lives is variable, with several factors impacting this number, e.g., race, gender, employment status, and family situation~\citep{Zhaoyang_PsyAging}. One important determinant of the number of social interactions in a typical day is the time left in life~\citep{Carstensen_AmPsy}.

To a certain extent, the number of interactions we have in our daily lives is a consequence of our present and past choices; that will depend, among other factors, on the challenges we face in our lives, in particular at work. If we have to perform complex tasks, it is necessary to be part of a team, and the size of the team will, in general, be related to the complexity of the task. In this sense, we characterize the present work as a continuation and a generalization of previous works where the number of players in a game is not fixed \emph{a priori}. We refer, in particular, to \citep{Souza_Pacheco_Santos:JTB2009,Archetti_JTB}, which studied the effect of the size of the groups on the outcome of given games. In~\citep{Izquierdo,Kurokawa_JTB19}, an individual opt-out strategy was introduced, having as a pratical consequence the variation of the size of the game. Symmetric games with variable number of players in neural networks also appear in~\citep{Gatchel_2021}. If the possible interactions in a population are modeled using a graph, games played by each individual might be variable as in~\citep{Broom_Rychtar2012,Erovenko_etal_2019}. Furthermore, the individual contact network may evolve according to the outcome of the game, and, therefore, the number of interactions in each game varies with time~\citep{TaylorNowak_2006,Skyrms_Pemantle_PNAS2006,Pacheco_2006}.

The objective of this work is to go beyond these previous works and to consider that the definition of the number of players that will participate in a given game is a player's decision. In \emph{variable size game theory} (VSGT), there is always a first step in which an individual is chosen from the population --- the so-called \emph{focal player} ---, and he or she will decide the number of players that will participate in the game. Only after that initial step will the group of players be selected and the game be played.

The outline of the paper is as follows: we introduce the basic notation in Sec.~\ref{sec:definition}. In Sec.~\ref{sec:examples}, we present several applications. In particular, in Subsec.~\ref{ssec:games}, we discuss the traditional structure of game theory: dominance, coordination, and coexistence, stressing the fact that a single change in a player's strategy, without changing the game's pay-off, may change game types. We also find the game with a fixed number of players that reproduces the fixation pattern of the given VSGT. In Subsec~\ref{ssec:eusociality}, we propose a simple model for the evolution of eusociality, introducing a game with prosocial (i.e., large $d$) and antisocial (i.e., small $d$) players, while in Subsec.~\ref{ssec:spacial} we show that a player strategy may provide a reason for the spontaneous separation of types, without a clear physical barrier, providing one possible scenario for speciation. In the last example, in Subsec.~\ref{ssec:compartimental}, we show how to derive classical differential equations from compartimental models using VSGT. Finally, in Sec.~\ref{sec:conclusion}, we discuss our findings, possible future work, and applications.

\section{Definitions}
\label{sec:definition}

Let us consider a fixed-size population, composed of two different types, say, \A and \B. We define a family of $d$-player symetric games, where $1\le d\le d_{\mathrm{max}}<\infty$, $d\in\N$. For any value of $d$, $\mathbf{a}^{(d)},\mathbf{b}^{(d)}\in\R^{d}$, where $a_k^{(d)}$ ($b_k^{(d)}$) indicates the pay-off of an \A-type (\B-type, respectively), in a $d$-player game, where $k$ individuals are of type \A.
If there are $i$ individuals of type \A in a population of $N$ individuals, pay-offs are defined by, respectively~\citep{GokhaleTraulsen_PNAS10,Lessard_DGA11},
\begin{align}
\label{eq:def_psiA}
\psi_\A^{(d)}(i)&=\sum_{k=0}^{d-1}\frac{\binom{i-1}{k}\binom{N-i}{d-1-k}}{\binom{N-1}{d-1}}a_k^{(d)}\ ,\\
\label{eq:def_psiB}
\psi_\B^{(d)}(i)&=\sum_{k=0}^{d-1}\frac{\binom{i}{k}\binom{N-i-1}{d-1-k}}{\binom{N-1}{d-1}}b_k^{(d)}\ .
\end{align}
In this case, we say that the population is at state $i$.
We assume that the $d$ players were drafted from the population with equal probability.
Note that $\psi_{\A,\B}^{(d)}$ are polynomials of degree $d-1$.

The first step in the proposed model is to select an individual in the population, with equal probability, to act as the \emph{focal player}. The focal player will decide the number of players that will participate in the game, according to the probability distributions
\[
 \bm{\lambda}_{\A}\bydef\left(\lambda_\A^{(d)}\right)_{d\le d_{\max}}\quad\text{or}\quad\bm{\lambda}_{\B}\bydef\left(\lambda_\B^{(d)}\right)_{d\le d_{\max}}\ ,
\]
if the focal player is of type \A or type \B, respectively. The probability distribution is such that $\lambda^{(d)}_{\A,\B}\ge 0$, and $\sum_{d=1}^{d_{\max}}\lambda^{(d)}_{\A,\B}=1$.

The pay-off of the focal player is given by $\Psi_{\A,\B}(i)\bydef\sum_{d=1}^{d_{\max}}\lambda_{\A,\B}^{(d)}\psi_{\A,\B}^{(d)}(i)$, for type \A and type \B individuals, respectively. For non-focal players, pay-offs are given by $\mu\sum_{d=1}^{d_{\max}}\lambda_{\mathbb{X}}^{(d)}\psi_{\A,\B}^{(d)}(i)$, where $\mathbb{X}=\A$ or \B indicates the focal player. The parameter $\mu\in[0,1]$ introduces an asymmetry between focal and non-focal players that is natural in the model, given a possibly reduced pay-off to non-focal players. In this work, we will consider only the case in which each strategist defines precisely the number of players to be involved in the game, i.e. $\lambda_{\A}^{(d)}=\delta_{d,d_{\A}}$ and $\lambda_{\B}^{(d)}=\delta_{d,d_{\B}}$, with $d_\A,d_\B\le d_{{\max}}$, and $\delta_{\cdot,\cdot,}$ is the Kronecker delta. In the examples studied, we will consider only $\mu=0$, which means that the focal player acts as group leader, dictator, or similar and $\mu=1$, in which case the game is symmetric.

\begin{remark}
Note that if $\bm{\lambda}_{\A}=\bm{\lambda}_{\B}$, the payoff of a player will not depend on the strategy of the focal type. As all players have the same probability of being focal, all pay-offs may be replaced by equivalent functions, and the parameter $\mu$ may be ignored. If the above equality does not hold, there are no simple equivalent pay-off functions, as it is necessary to consider that the probability that a given individual will be selected to participate in a game as a non-focal player will depend on the strategy of both players and the population state.
\end{remark}

From now on, we will use also the expression \emph{absolute fitness} or simply \emph{fitness}, to refer to the average pay-off of the individuals of a given type, once all individuals in the population have had the opportunity to be the focal player. In this sense, the pay-off is a game property, while fitness is a state property. The relative fitness $\Theta(i)$ is the ratio between the fitness of type \A and \B at state $i$. When $\mu=0$, this equals 
\begin{equation}\label{eq:relative_fitness}
\Theta(i)=\frac{\Psi_{\A}(i)}{\Psi_{\B}(i)}\ .
\end{equation}
We define the \emph{fitness potential}
\begin{equation}\label{eq:potential}
V\left(\frac{i}{N}\right)=-\sum_{j=0}^i\log\Theta(j)\ ,
\end{equation}
which provides simple heuristics to analyze the game dynamics, cf.~\citep{ChalubSouza_JTB18,ChalubSouza_JMB16}.
In this work, we will only refer to the fitness potential if $\mu=0$.

\begin{remark}\label{rmk:wsp}
Assume $\mu=0$. Let us define $x=i/N$, and assume the \emph{weak selection principle}
\begin{equation}\label{eq:weak_fitness}
\psi_{\A,\B}^{(d)}(x)= 1+\frac{1}{N}\varphi_{\A,\B}^{(d)}(x)+o(N^{-2})\ ,
\end{equation}
for large values of $N$, uniformly on $x$. We use the ``small o'' notation to indicate the asymptotic behavior of relevant variables in the limit $N\to\infty$. We define $\Phi_{\A,\B}(x)\bydef\sum_{d=1}^{d_{\max}}\lambda_{\A,\B}^{(d)}\varphi_{\A,\B}^{(d)}(x)$. Therefore,
\begin{align*}
V(x)&=-\sum_{j=0}^i\log\frac{\Psi_{\A}(x)}{\Psi_{\B}(x)}\\
&=-\sum_{j=0}^i\log\frac{1+\frac{1}{N}\Phi_{\A}(x)+o(N^{-2})}{1+\frac{1}{N}\Phi_{\B}(x)+o(N^{-2})}\\
&=-\sum_{j=0}^i\left\{\frac{1}{N}\left(\Phi_{\A}(x)-\Phi_{\B}(x)\right)+o(N^{-2})\right\}\\
&=-\int_0^x\left(\Phi_{\A}(x)-\Phi_{\B}(x)\right)\rd x+o(N^{-1})\ .
\end{align*}
As discussed in~\citep{ChalubSouza14a,ChalubSouza_JTB18}, the fixation probability $\phi(x)$ is the stationary solution of the adjoint Kimura equation with appropriate boundary conditions; namely
\[
 \phi(x)=\frac{\int_0^x\e^{\frac{2}{\kappa}V(y)}\rd y}{\int_0^1\e^{\frac{2}{\kappa}V(y)}\rd y}\ ,
\]
where $\kappa$ is the selection strength (inverse population size).
This result is valid in the limit $N\to\infty$ for fitnesses/pay-off functions given by Eq.~(\ref{eq:weak_fitness}). In this case, type \A individuals will be the most likely to fixate if the boundary $x=1$ can be reached with less (fitness potential) energy than the boundary $x=0$;  otherwise, type \B will be the most likely to fixate. If both boundaries require a comparable amount of energy to be reached, there will be a metastable coexistence state and fixation will take longer. For further details, see~\citep{ChalubSouza_JTB18}. In this work, except when otherwise stated, we will depart from the weak selection regime,  cf.~\citep{ChalubSouza_JMB16}.
\end{remark}

A differential equation assumes at least implicitly an infinite population; in part of the examples described in the present work, we will consider finite populations, which are compatible with infinite ones if we assume the weak selection principle as described above.

Throughout this work, we will always consider the Wright-Fisher (WF) model for finite populations~\citep{Fisher_1922,Wright_1931}. The WF model is characterized by a $(N+1)\times(N+1)$ stochastic transition matrix $\mathbf{M}=\left(M_{ij}\right)_{i,j=0,\dots.N}$, where $N$ is the population size and $M_{ij}=\binom{N}{i}\tsp_j^j(1-\tsp_j)^{N-i}$ is the transition probability from state $j$ to $i$. The type selection probability $\bm{\tsp}\bydef(\tsp_0,\tsp_1,\dots,\tsp_N)$ was introduced in~\citep{ChalubSouza_JMB17} and is related to the fitness by
\[
\tsp_i=\frac{i\Psi_\A(i)}{i\Psi_\A(i)+(N-i)\Psi_\B(i)}\ .
\]
We assume no mutation in the model, i.e., $s_0=0$, $s_N=1$.

The fixation probability vector $\mathbf{F}=(F_0,\dots, F_N)$ is obtained considering the Wright-Fisher process. Let $F_i$ be the probability that type \A is fixated in the population, starting from a presence of $i=0,\dots,N$ type \A individuals, i.e., the fixation probability from initial condition $i$ is equal to to the sum over all $j$ of the probabilities that, in the first step, the state changes from $i$ to $j$, and then it fixates from $j$, $F_i=\sum_jF_jM_{ji}$. Then, $\mathbf{F}$ is the only solution of the eigenvalue problem $\mathbf{F}^\dagger\mathbf{M}=\mathbf{F}^\dagger$, $F_0=1-F_N=0$. Namely, we define $\widetilde{\mathbf{M}}=\left(M_{ij}\right)_{i,j=1,\dots,N-1}$, $\mathbf{b}=\left(M_{Ni}\right)_{i=1,\dots,N-1}$, $\widetilde{\mathbf{F}}=\left(F_i\right)_{i=1,\dots,N-1}$ and then $\widetilde{\mathbf{F}}^\dagger=\mathbf{b}^\dagger\left(\mathbf{I}_{N-1}-\widetilde{\mathbf{M}}\right)^{-1}$, where $\mathbf{I}_N$ is the $N\times N$ identity matrix. In the above formulas, vectors are column-vectors and $(\cdot)^\dagger$ denotes their transpose. See~\citep{ChalubSouza_JMB17} for further details.

\begin{remark}
Other finite population models (e.g., the Moran process~\citep{Moran_1962}) could be used as well. In this work, we will only use the WF process for two main reasons. The first one is that the reproductive and the Darwinian fitnesses are compatible for all choices of parameters. A second reason is that the inverse problem (to be discussed in Subsec.~\ref{ssec:games}) can be applied to any fixation function --- it is important to note that in $d$-player game theory, for $d>2$, the fixation probability is not necessarily increasing in. For further details, see~\citep{ChalubSouza_JMB17,ChalubSouza_BMB19}.
\end{remark}

\section{Examples}
\label{sec:examples}

\subsection{Dominance, coexistence, and coordination}
\label{ssec:games}

The replicator equation for a two-type population is given by 
\[
x'=x(1-x)\Delta\psi(x)\ ,
\] 
where $\Delta\psi\bydef\psi^{\A}-\psi^{\B}$ is the fitness differences between the types; if $\Delta\psi(x)$ does not change sign, we say that \A dominates \B (if the difference is positive) and that \B dominates \A otherwise. Note that $x(t)\stackrel{t\to\infty}{\longrightarrow} 1$ in the first case and $x(t)\stackrel{t\to\infty}{\longrightarrow} 0$ otherwise for all non-trivial initial conditions. If the same fitness difference is used to model the Wright-Fisher process, the fixation probability will be close to 1 in the first case and close to 0 in the second, except if the initial condition is near the boundaries. In the following discussion, we will ignore cases in which $x(0)$ is close to 0 or 1. We also assume that $\Delta \psi$ is smooth.

If $\Delta\psi(x)>0$ for $x< x_*\in(0,1)$ and $\Delta\psi(x)<0$ for $x>x_*$, then $x(t)\stackrel{t\to\infty}{\longrightarrow}x_*$, and the fixation probability of the Wright-Fisher dynamics is characterized by a plateau, apart from the boundaries. This is characteristic of a loss of memory of the initial condition, as initially, the system converges to an inner equilibrium at $x_*$. In the finite population case, the system will fixate or will be extinct with comparable probabilities. In the replicator dynamics, $x_*$ is an inner stable equilibrium. This is the coexistence case.

On the other hand, if  $\Delta\psi(x)<0$ for $x< x_*\in(0,1)$ and $\Delta\psi(x)>0$ for $x>x_*$, then $x(t)\stackrel{t\to\infty}{\longrightarrow} 0$ or $x(t)\stackrel{t\to\infty}{\longrightarrow}1$ according to the case $x(0)<x_*$, $x(0)>x_*$, respectively. The fixation probability is close to 0 for $x<x_*$ and close to 1 for $x>x_*$, with a sharp discontinuity around $x_*$. This is the coordination case.

For a $2$-player game, this exhausts all possibilities as $\Delta\psi$ is an affine function. The analysis is simple if we study the potential defined in the weak selection approximation $V$, which is a quadratic function. An inner maximum (minimum) indicates an unstable (stable, respect.) equilibrium in the replicator equation, and a sharp discontinuity (plateau, respect.) in the fixation probability of the Wright-Fisher process. For $d$-player games, $\Delta\psi$ is a polynomial of degree $d-1$ and the situation is more complex, with sharp discontinuities and plateaux coexisting in the same fixation function (as inner stable and unstable equilibria will be present in the associated replicator dynamics), cf.~\citep{ChalubSouza_JMB16,ChalubSouza_JTB18}.

In the discussion above, it is clear that the fixation pattern will depend on the function $\Delta\psi$ and, therefore, only on the pay-offs of the game. However, the situation is different in VSGT, as we have families of games, indexed by $d$ with possibly different classifications. Therefore, without changing the family of pay-offs, but changing the game size preferences of the players, we may change from one pattern to a completely different one.

On the other hand, in a previous work of one of the authors, it was proved that any fixation pattern can be arbitrarily approximated by the Wright-Fisher dynamics, with pay-offs given by a certain $d$-player game, if $d$ is large enough. To obtain the minimum $d$ that approximates a given fixation pattern within an acceptable approximation, it is necessary to solve an inverse problem, cf.~\citep{ChalubSouza_BMB19}. It is clear that the equivalent fixed-size game will be strongly dependent not only on the family of pay-offs, but also on the individual strategies with respect to game sizes.

\begin{figure*}[htb]
\centering
\includegraphics[width=\textwidth]{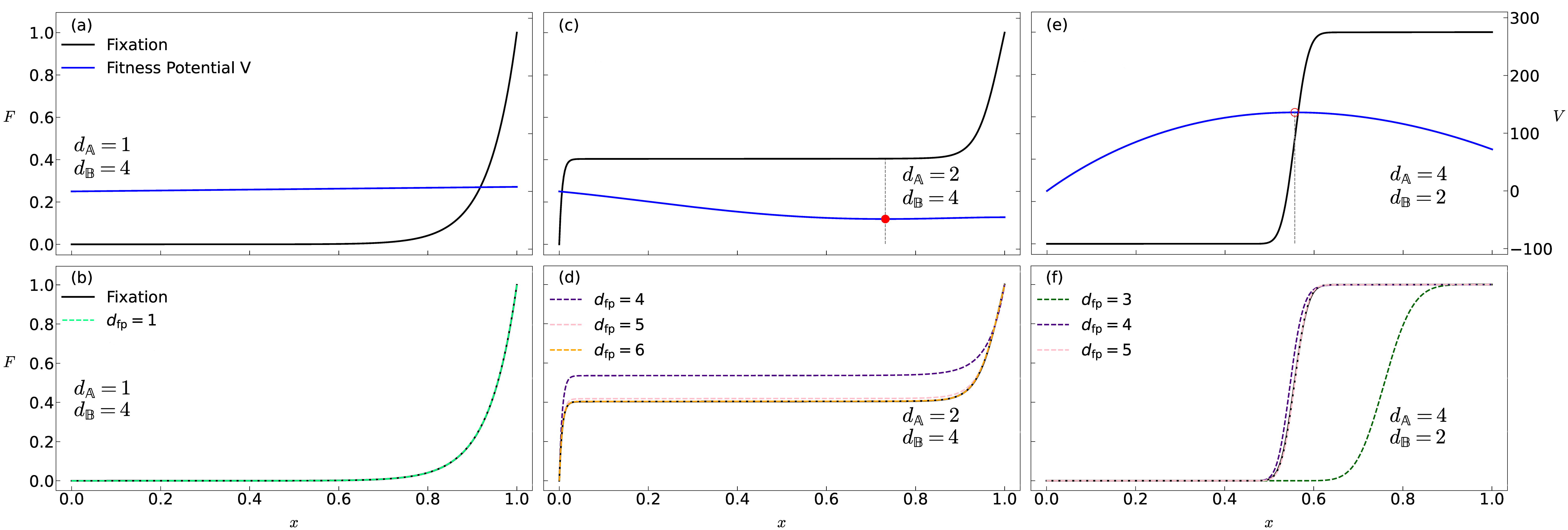}
\caption{(a) For any $d$-player game, the pay-off of \B individuals is larger, $a_k^{(d)}=1$, $b_k^{(d)}=1.02$ for all $k$ and $d\le d_{\max}=4$. We assume that type \A tries to avoid the game, while \B individuals want games as large as possible, i.e.,  $d_\A=1$, $d_\B=4$. The equivalent fixed-player game is presented in subfigure (b), with $d_{\fp}=1$ and pay-offs given by $\ba^{(1)}_\fp=(93.07)$, $\bb^{(1)}_\fp=(94.94)$. Note that all pay-off may be multiplied by a constant without changing the game's result. In this case, $\bb^{(1)}_\fp=1.02\times\ba^{(1)}_\fp$. In subfigures (c) and (d), we consider the direct and inverse problem, respectively, for $\ba=\left(\ba^{(1)},\dots,\ba^{(4)}\right)=\left((1),(6,2),(1,1,1),(1,2,3,2)\right)$ and $\bb=\left(\bb^{(1)},\dots,\bb^{(4)}\right)=\left((1),(4,1),(1,1,1),(5,3,4,2)\right)$, i.e., \A dominates the 2-player game, \B dominates the 4-player game and the others are neutral. We assume $d_\A=2$ and $d_\B=4$. The pay-offs are chosen such that every time an \A or \B player is selected as focal, his or her fitness will increase more than it will decrease in case his or her opponent is selected. The resulting game is of coexistence type and the equivalent fixed-player game requires at least 6 players in each game with $\ba^{(6)}_\fp=(0.0160, -0.0425, -0.159,  5.38, -0.982,   0.905)$ and $\bb^{(6)}_\fp=(0.0138, -0.0322, -0.221,  5.37, -0.839,  0.883)$. In subfigures (e) and (f), we consider the same pay-off as (c) and (d), but assume that $d_\A=4$ and $d_\B=2$, i.e., each player chooses (in a non-rational way) a game in which he or she is at a disadvantage. The resulting game is equivalent to a coordination type game, with $d_\fp=5$ and $\ba^{(5)}_\fp=(1.03, 0.849, -0.335, -0.194,  0.501)$, $\bb^{(5)}_\fp=(4.16, -0.367,  0.0973, -0.331,  0.531)$. In the last two cases we also plot the fixation probability of the game with higher accuracy for suboptimal $d_\fp$, with consistent colors in subfigures (d) and (f). In the three figures above, fixation probabilities are represented by continuous black lines, while fitness potentials are given by continuous blue lines. In the three pictures below, fixation probabilities are also denoted by continuous black lines, but they are superimposed on the fixation of the equivalent $d$-player game.}\label{fig:games}
\end{figure*}

More precisely, in this subsection, we consider a population of two types of individuals which evolves according to a variable $d$-player game, in which $d_{\max}=4$. We assume that only the focal player receives the pay-off, i.e., $\mu=0$. The relative fitness is given by~Eq.~\eqref{eq:relative_fitness}. The fixation probability is obtained by solving the associated Wright-Fisher process, as explained in the introduction, and in the sequel we consider the inverse problem, discussed in~\citep{ChalubSouza_BMB19}, to obtain a symmetric game with a fixed number of players $d=d_\fp$ that reproduces with accuracy the fixation probability; in this case, we say that the game of fixed size is \emph{equivalent} to the game of variable size proposed. In Fig.~\ref{fig:games} we explore the three traditional cases studied in game theory, dominance, coexistence, and coordination~\citep{HofbauerSigmund}. In the dominance case, we considered $d$ and $k$ independent pay-offs, such that $b_k^{(d)}>a_k^{(d)}$, and $d_\B>d_\A=1$. In the long run, the population will fixate in the \B type, but as the pay-offs are independent of the values of $k$ and $d$, the fixation probability can be obtained from an equivalent 1-player game, i.e., the constant fitness case.

We also show that we can move from the typical fixation of a coexistence-like game to a coordination-like game without changing the payoff. In the former case, there is an essentially constant plateau for the fixation probability, for initial presence $x$ not too close to 0 (extinction) nor 1 (fixation). The plateau indicates the existence of an inner equilibrium of the replicator equation. In the latter case, there is a jump from 0 to 1 in the fixation probability at $0\ll x_*\ll1$. We showed that we could move from one case to the other changing only the strategic decision of type \A and type \B individuals with respect to the choice of the number of players in each given game.

\begin{remark}\label{rmk:one}
Note that the definition of the potential $V$ in Eq.~(\ref{eq:potential}) requires the value of $\Theta(0)$ and $\Theta(N)$. Therefore, the values of $\Psi^{(\A)}(0)$ and $\Psi^{(\B)}(N)$, and, consequently the definition of the pay-off functions of a given type, when that given type is absent of the population, i.e., $\psi^{(d)}_{\A}(0)$ and $\psi^{(d)}_{\B}(N)$, must be defined for all values of $d$ such that $\lambda^{(d)}_{\A}\ne0$, or $\lambda^{(d)}_{\B}\ne0$, respectively. On the other hand, 
\[
\binom{-1}{k}=\frac{1}{k!}\prod_{i=0}^{k-1}(-1-i)=\frac{(-1)^k}{k!}\prod_{i=1}^ki=(-1)^k
\]
is the only natural extension of the binomial function, a fact explicitly used to define $\psi^{(d)}_{\A}(0)$ and $\psi^{(d)}_{\B}(N)$ from Eqs.~(\ref{eq:def_psiA}) and (\ref{eq:def_psiB}), and, therefore to correctly define $V$ from Eqs.~(\ref{eq:relative_fitness}) and~(\ref{eq:potential}).
\end{remark}

\subsection{Evolution of eusociality}
\label{ssec:eusociality}

According to~\citet{smith1997major}, a group of individuals is \emph{eusocial} if it contains sterile workers that help their parents; typical examples are insects, but this is also the case of the naked mole rat, a mammal. However, in~\citep{NowakTarnitaWilson_Nature2010}, the necessity to consider relatedness between individuals in the same group was questioned, challenging the foundations of the inclusive fitness theory. This idea generated a large discussion in the specialized literature, cf., e.g. ~\citep{FerrierMichod_Nature2011,HerreWcislo_Nature2011} and the answer in~\citep{NowakAllen_PLOS2015}. Here we explore a possible explanation for the evolution of eusocial behavior within the framework of VSGT, and therefore, relatedness among individuals is not required.

We define the eusocial behavior as a given individual in a population working for the benefit of a different, not necessarily genetically related, individual. In the proposed model, we consider two different types of players, that decide to work or not to the benefit of the focal player: type \A players have the cognitive capacity to gather larger groups of individuals and collaborate in the pursuit of complex tasks; on the other hand, type \B individuals are unable to gather larger groups and even if he or she joins a group large enough to fulfill the task, will be unable to collaborate or is not willing to do so. The result of the task will benefit only the focal player; considering successive rounds of the game, every individual will have the opportunity to be the focal player and get the lion's share of the completed task.

More precisely, we consider that $k_{\min}$ eusocial (type \A) individuals are required to fulfill the task, which gives an increase in $G>0$ in the absolute fitness of the focal player, which also pays a cost $c>0$ if he or she is of type $\A$, and pays nothing otherwise. For all the other players, the pay-off is identically zero, independently of the fact that the task has been completed or not. More precisely, $0<a_k^{(d)}=b_k^{(d)}=\varepsilon$, for $k<k_{\min}$ (where $\varepsilon$ prevents the existence of zero entries in the transition matrix and is given to all individuals in the population regardless of their participation or strategy). If $k\ge k_{\min}$, pay-off's are given by $a_k^{(d)}=\varepsilon+G-c$ and $b_k^{(d)}=\varepsilon+G$. We assume $\mu=0$, and therefore only the focal player receives the pay-off. The pay-offs of other players are identically zero. In particular, participation in the pursuit of the task is evolutionarily neutral to non-focal individuals, stressing that the model puts all focus on the strategic definition of the group size.

If there is only a small number of type \A individuals, evolution will be locally neutral, as no group will have the required number of eusocial individuals. If we assume $d_{\A}>k_{\min}>d_{\B}$, the same will happen whenever a type \B individual is selected as a focal individual. If the number of type \A individuals is sufficiently large when compared to $k_{\min}$, the presence of type \A individuals may increase randomly such that it is possible to fulfill the task.

\begin{figure*}
\centering
\includegraphics[width=\linewidth]{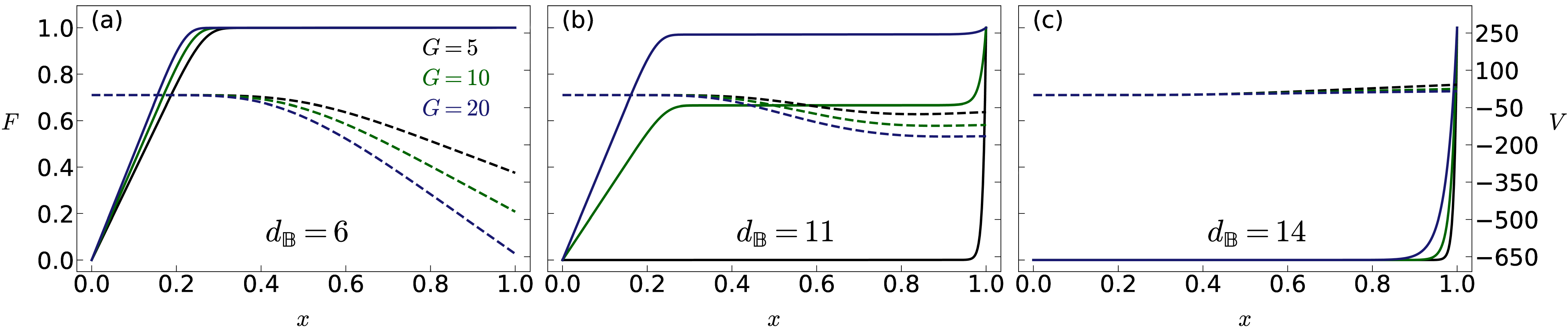}
\caption{Consider the model for the evolution of eusociality, with gain $G=5, 10, 20$ (marked with different colors), cost $c=1$, $\varepsilon=1$, $\mu=0$, minimum size group $k_{\min}=8$, and $d_\A=14$. The fixation probability is obtained from the Wright-Fisher process with population $N=400$ (continuous lines, scale to the left). Fitness potentials are  indicated by dashed lines of the same color, with values to the right of the graph. We consider three cases, $6=d_\B<k_{\min}$ (left), where prosocial individuals will prevail if their initial presence is large enough; $14=d_\B=d_\A$ (right), where prosocial individuals will be eliminated by natural evolution (except if their initial presence is close to 1, in which case they can prevail due to stochasticity); and $k_{\min}<11=d_\B<d_\A$, a transition regime, where the viability of the prosocial individual is uncertain. In the last case, natural selection will lead to a coexistence regime, associated with a local minimum of the potential $x_*\in(0,1)$ and the final state will be defined by stochasticity, cf. the discussion at~\citep{ChalubSouza_JMB16,ChalubSouza_JTB18}.}
\label{fig:PGG}
\end{figure*}

If $d_{\A}>k_{\min}>d_{\B}$, and the initial fraction of type \A individuals is smaller than a certain critical value, the result of evolution is uncertain; after that critical value, the long-term dominance by eusocial individuals is almost certain. This follows from the fact that \B-type individuals are unable to gather sufficiently large groups to perform the task. When $d_\B=d_\A$, type \B has similar skills as type \A individuals in forming larger groups and, therefore, will dominate in the long run. The eusocial behavior will not prevail, and the same is true if $d_\B>d_\A$. If $k_{\min}<d_\B<d_\A$, type \B individuals are able to gather large enough groups, but the presence of type \A individuals in these groups (i.e., that the group will have a sufficient amount of prosocial individuals to perform the task) is not guaranteed. Therefore, this is a transition region where the result will present a strong dependence on the value of $G$; a larger value of $G$ will favor prosocial behavior. See Fig.~\ref{fig:PGG}.

\subsection{Spontaneous geographic separation}
\label{ssec:spacial}
Following ideas from~\citet{Nagylaki_1992}, we use the Island Model to study spontaneous separation, according to the game size that each type plays. In the proposed model, contrary to what may be suggested by the terminology, we discuss a possible geographic separation, but without assuming any physical barrier between different environments, i.e., there is no spatial segregation among types. We consider a model of a small island, with population $N_{\mathrm{i}}$,  close to a large continent with population $N_{\mathrm{c}}\gg N_{\mathrm{i}}$. These numbers also correspond to the carrying capacities of both environments, and the population will be continuously adjusted such that after each interaction the real population size is equal to the carrying capacity. We refer to the island and the continent as \emph{patches}, and the name only reflects the difference in population size; in particular we allow free migration between both patches, depending on the individual's strategy.

The population consists of two types of individuals, \A and \B, that are similar with respect to all characteristics, except regarding game size preferences $d_{\A}$ and $d_{\B}$. Type $\A$ likes gathering large groups, while \B prefer small groups, i.e., $d_{\A}>d_{\B}$. This is similar to the model developed in Subsec.~\ref{ssec:eusociality}, but in this case, there is no task to be performed. Both types are initially identically distributed (50\%-50\%) in both patches. Each type feels comfortable in an environment if they meet in the game partners with a certain amount of players of identical type $k_{\min}$, namely $a^{(d)}_k=1$ for $k\ge k_{\min}$, and $0$ otherwise, and $b^{(d)}_k=1$ for $k<d-k_{\min}$ and 0 otherwise. To stress that all the conclusions follow solely from the game size that each type plays, $k_{\min}$ will be the same independently of the focal player. The pay-off is converted, uniquely to the focal player (i.e., $\mu=0$), into the willingness to stay in the same patch (if the fitness is $1$) or to migrate to a different one with probability $p\in(0,1]$, if the fitness is 0.

After each individual in the population has played as a focal player, we update the population in both patches, and, consequently, the population size in each patch will be different from the carrying capacity. If it exceeds it, the exceeding number is removed with an equal probability for all individuals in the given patch. If the population is below the carrying capacity, then individuals are randomly selected to reproduce, with an equal probability for all individuals in the given patch, until the carrying capacity is reached.

\begin{figure*}[htb]
\includegraphics[width=\textwidth]{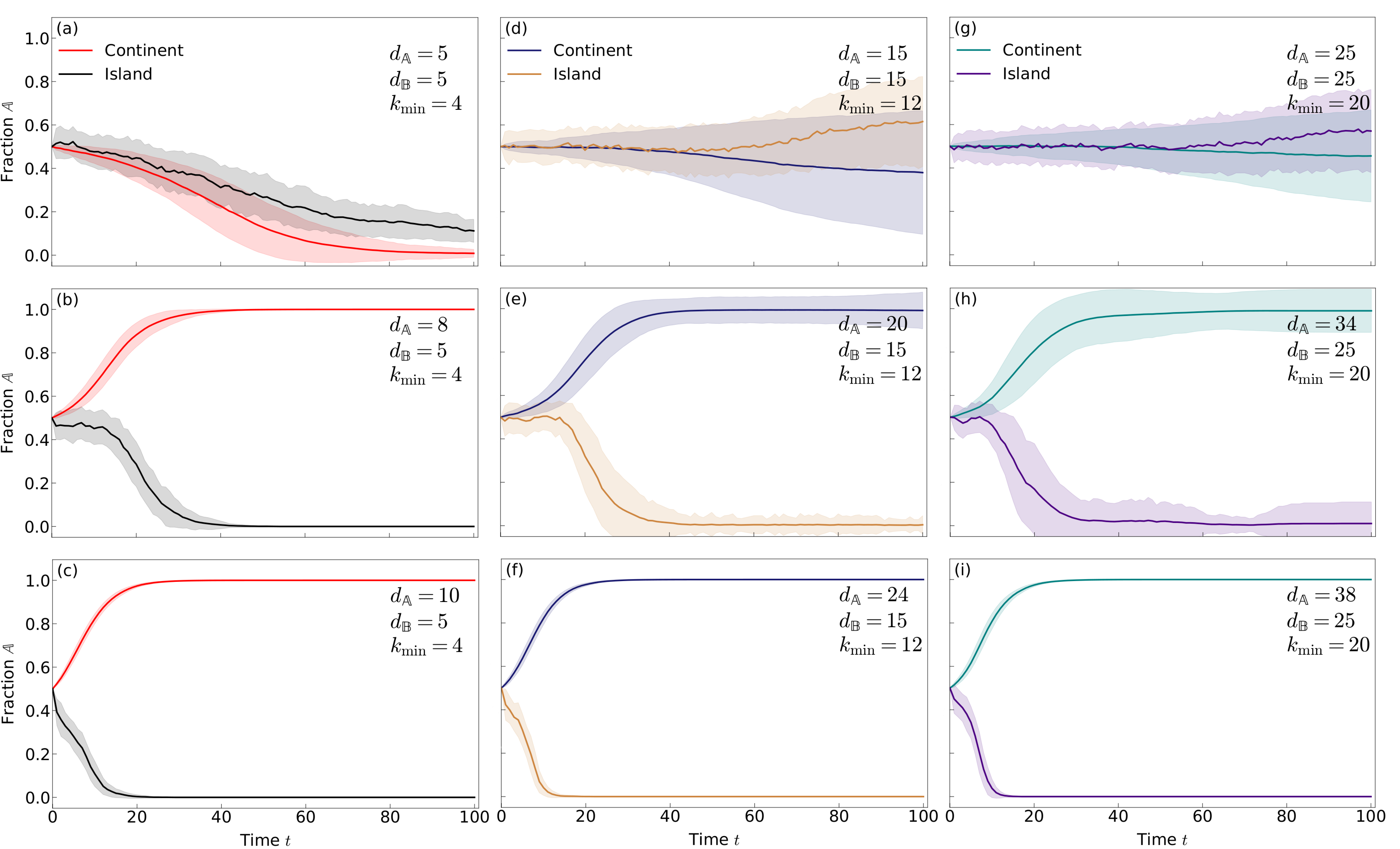}
\caption{We consider three sets of simulations of the continent-island model. In the first row, we consider $d_\A=d_\B=5,15,25$ (left, middle, right, respectively), with $k_{\min}=4, 12, 20$, always smaller than $d_{\B}$. Evolution is ``quasi-neutral'' and, after 100-time steps, there is no clear spatial separation. Note that, in each one-time step, all individuals in both patches have the opportunity to migrate, according to the result of the game. For $d_{\A}>d_{\B}$ (second and third rows), the separation is clear, with the continent dominated by type-\A individuals and the island by type-\B individuals. Values for $d_\A$, $d_\B$ and $k_{\min}$ are indicated in the figures. The size of the population of the island is $N_{\mathrm{i}}=50$ and the continent is $N_{\mathrm{c}}=1000$, initial conditions are 25 and 500 individuals of each type in each patch, and the migrations probability is given by $p=0.2$. Error bars (shadows) are calculated after 100 simulations.}
\label{fig:separation}
\end{figure*}

The idea of considering migration as a result of a given game pay-off appears in~\citep{Broom_Rychtar2012,Erovenko_etal_2019}, where the authors consider the spatial structure of individuals in a given population modeled through a graph, allowing games played between groups of different size and analyzing the influence of different graph topologies. In this subsection, we use VSGT to model spontaneous type separation according to the strategy. This idea is natural, as the typical structure of the game is of coordination type, i.e., each type has a high fitness if his or her type predominates among game players. In fact, in our model, this is not exactly true, as $k_{\min}$ can be small when compared to $d_{\B}$. The more interesting phenomenon is not the separation, but the fact that individuals who prefer to play in large groups (i.e., type \A) individuals concentrate in the continent, with its larger populations, while type \B individuals move to the island, mimicking migration from rural (\emph{island}) to urban (\emph{continent}) areas and vice-versa. The separation will be faster depending on the absolute value of $d_{\A}-d_{\B}$ and on $p$. In the limit case in which $d_{\A}=d_{\B}$, evolution is purely random, stressing the fact that the spontaneous separation is due uniquely to the strategic difference in game size, see Fig.~\ref{fig:separation}.

Finally, we would like to stress that the above model resembles the concept of \emph{sympatric speciation}. Speciation occurs when a single species splits into two different species. There are several evolutionary mechanisms possibly involved in the speciation process. The most relevant one is  \emph{allopatric speciation} when a geographical (physical) barrier provides segregation of the parent species into two groups with small or non-existent gene flow among them. We note that in our model there is no physical barrier between the island and the continent (despite the choice of words), and every individual is free to migrate but decides not to do so if he or she is in an environment of similar individuals. If, in the long run, we identify the types with newly formed species (due to accumulated mutation in possibly different loci), then we have an example of \emph{sympatric speciation}, cf.~\citep{Gavrilets+2004,ErnestMayr}.

\subsection{Compartimental models}
\label{ssec:compartimental}

In this section, we show how to derive the SIRS epidemic model as a VSPG game.
This requires a level of generalization of the theory discussed before, as in this case each individual has more than two possible strategies, each one corresponding to the possible compartments in the model.

We will assume a symmetric game, i.e., with $\mu=1$.
Consider a population of $N$ individuals, of three types,  \textbf{S}uscpetible, \textbf{I}nfectious, and \textbf{R}ecovered. The number of individuals in each class will be given by $S$, $I$, and $R$, respectively. Let $N\bydef S+I+R$ be the total population.

We compute payoffs according to the structure of the SIRS model, namely $S+I\stackrel{\beta}{\longrightarrow} I+I$, $I\stackrel{\gamma}{\longrightarrow} R$, $R\stackrel{\alpha}{\longrightarrow} S$. Note that in our parlance, the first transition will be represented by a two-player game, while the last two as one-player games. In this model, all participants receive the payoffs and not only the focal player.

\begin{figure*}[htb]
\centering
\includegraphics[width=\textwidth]{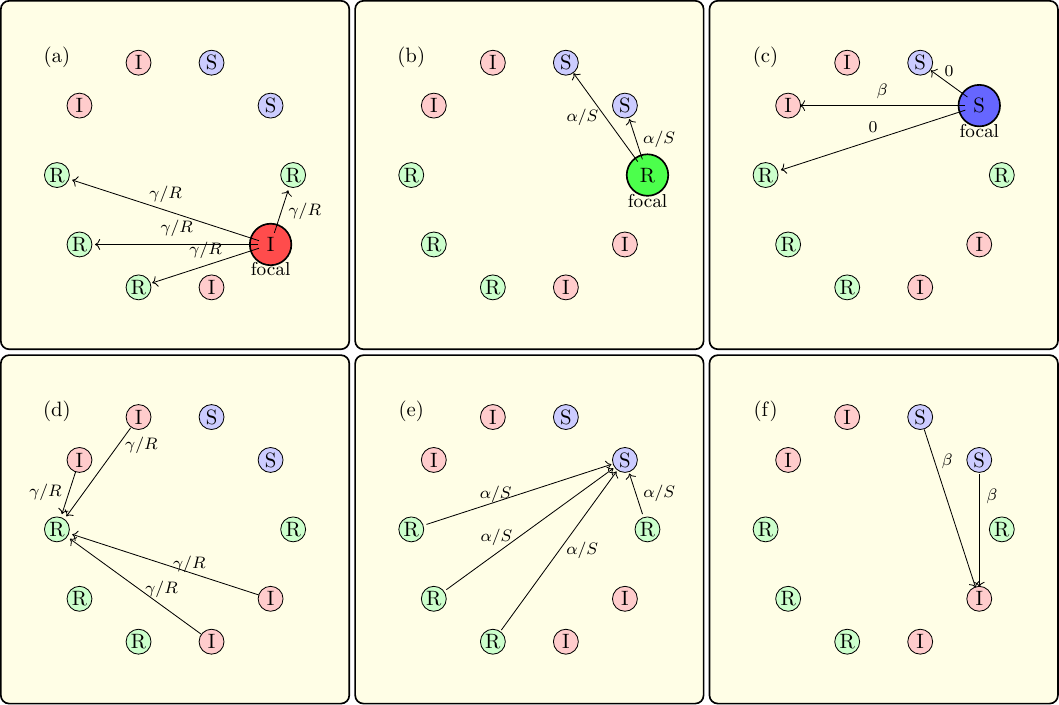}
\caption{Consider a population of $N=10$ individuals of one of the three types: \textbf{S}, \textbf{I}, and \textbf{R}. The three figures above show the fitness loss from the focal individual to all individuals in the population, while the row below shows the fitness gain of an individual from all the other individuals in a given game. If the individual selected as the focal player is a type \textbf{I} individual, its fitness is decreased by $\gamma$, distributed among all \textbf{R} individuals (a). The fitness of an \textbf{R} individual will be increased by $\gamma/R$ times $I$, the number of $I$ individuals in the population (d). If the focal type is of type \textbf{R}, its fitness is decreased by $\alpha$ (b), and the fitness of type \textbf{S} is increased by $\alpha/S$ for each \textbf{R} present in the population (d). Finally, if the focal individual is of type \textbf{S}, it is necessary to select a second individual. If the second individual is of type \textbf{I} (what happens with probability $I/N$), then the \textbf{S} individual loses a fitness $\beta$ and nothing happens otherwise (c). On the other hand, an \textbf{I} individual receives a fitness $\beta$ from an \textbf{S} individual \emph{if he or she was selected as the second player}, what happens with probability $1/N$ (f). All non-represented arrows mean a fitness transfer of 0; cf. Eqs~(\ref{eq:transitionS}), (\ref{eq:transitionI}), and (\ref{eq:transitionR}).}
\label{fig:SIRmodel}
\end{figure*}

Consider a given $\textbf{S}$ individual. If he or she was selected as the focal player, then a two-player game is played and his or her payoff will be changed only if the second player is of type \textbf{I}. His or her payoff will also be changed if the focal player is \textbf{R}, as \textbf{R} will \emph{donate} part of his or her payoff to the entire pool of \textbf{S} players (transition $R\stackrel{\alpha}{\longrightarrow}S$). Namely
\begin{equation}\label{eq:transitionS}
\Psi_{\textbf{S}}(S,I,R)=-\beta \frac{I}{N}+\alpha\frac{R}{S}\ .
\end{equation}
Now, consider an \textbf{I} individual. If he or she is the focal player, the pay-off decreases by a constant, representing the transition $I\stackrel{\gamma}{\longrightarrow} R$ (one-player game). The payoff of an \textbf{I} individual is also changed if the focal player was of \textbf{S} type and he or she was selected as the second player in the two-player game, what happens with probability $1/N$. Therefore
\begin{equation}\label{eq:transitionI}
\Psi_{\textbf{I}}(S,I,R)=\beta \frac{S}{N}-\gamma\ .
\end{equation}
Finally, the pay-off of an \textbf{R} individual decreases if it is focal and increases if \textbf{I} is focal, i.e.,
\begin{equation}\label{eq:transitionR}
\Psi_{\textbf{R}}(S,I,R)=\gamma\frac{I}{R}-\alpha.
\end{equation}
See also Fig.~\ref{fig:SIRmodel} for further information in the how Eqs.~(\ref{eq:transitionS})--(\ref{eq:transitionR}) were obtained.

It is clear that the average fitness is zero, i.e.,
\begin{align*}
&\overline{\Psi}(S,I,R)\\
&\quad\bydef S\Psi_\textbf{S}(S,I,R)+I\Psi_\textbf{I}(S,I,R)+R\Psi_\textbf{R}(S,I,R)\\
&\quad=0\ .
\end{align*}
The replicator equation is exactly the SIRS model:
\begin{align*}
S'&=S\left(\Psi_S(S,I,R)-\overline{\Psi}(S,I,R)\right)=-\frac{\beta S I}{N}+\alpha R\ ,\\
I'&=I\left(\Psi_I(S,I,R)-\overline{\Psi}(S,I,R)\right)=\frac{\beta S I}{N}-\gamma I\ ,\\
R'&=R\left(\Psi_R(S,I,R)-\overline{\Psi}(S,I,R)\right)=\gamma I-\alpha R\ .
\end{align*}

\section{Discussion}
\label{sec:conclusion}

The main novelty of the present work is to consider that games inside a population may vary in the number of participants as a consequence of a player's strategy.

We started by showing that the traditional examples of game theory can be recast into the framework of VSGT, and it is not necessary to change pay-offs to move from coexistence- to coordination-type games and vice-versa. A change in the prosociality (i.e., the typical number of participants in a game, when we are allowed to choose the game size) of the players involved in the game can have dramatic consequences in the long run.

As a second example, we showed that this framework is compatible with the fixation of a strategy in which individuals work for the benefit of a single one. The focal player is defined at random, which seems artificial if we think of a population of individuals, but natural when we consider that the evolving unit is the gene. The loosely defined \emph{eusocial gene} will be present in all individuals of a given type, but cannot know (abusing the language) which of its copies will manifest at the leader.

As a third example, in Subsec.~\ref{ssec:spacial}, we showed how a difference in the size of the game each type plays, along with the preference to be with similar individuals, may provide spontaneous geographical separation between types, even in the absence of a rigid barrier between different environment. This makes prosocial individuals migrate to the larger environment, while antisocial individuals move toward regions with smaller populations, in a phenomenon that resembles sympatric speciation.

As a last example, we also showed how to reformulate an epidemic model as a game in which transmission is seen as a two-player game and the other transitions, that do not depend on interactions in the populations, are considered as a one-player game. Each individual may play a two-player game if the focal player is of the Susceptible type and selects this individual -- possibly against his or her will -- to be part of the game.

In this work, we do not consider certain classical topics in game theory, for example, proofs of existence and classification of Nash equilibria, or the study of evolutionarily stable strategies. The proof of the existence of Nash equilibria follows without change from the one used in games of fixed size, as they are based only on the continuity of the pay-off function and the convexity and compactness of the domain of possible strategies (i.e., $S^{n-1}$ is a convex, bounded, and closed subset of $\R^n$). Note that in this case we probably will have to resort to mixed strategy games, defined, in this case, as non-trivial probability distributions $\bm{\lambda}$. The details will be developed in a follow-up work, where we also plan to study Evolutionarily Stable Strategies. One interesting question is the relation between these traditional concepts in game theory and the stability of the disease-free and endemic equilibria in epidemic models.

Many applications of these ideas can be expected in the near future. The most obvious application is in economic modeling, where the focal player may deliberately define the size of the game. Consider, for example, a hierarchical structure, as a company (or an academic department), in which all employees have to deal with one boss, but the boss has to deal with dozens, or more, employees, that can be or not be structured in smaller groups. Our framework may be used for the design of an optimal hierarchical structure. Another idea is to study how strategies vary over time, modeling the discussion present in~\citep{Carstensen_AmPsy}, in which when the number of expected future interactions decrease --- as a consequence of aging, for example --- more priority is given to interactions with affective value, i.e., with smaller $d$ but larger pay-offs. Note that the strategy-dependence on future expected interaction is a well-established concept in evolutionary game theory, following the seminal work by~\citet{Axelrod_Hamilton}; see also~\citep{Axelrod84}.

In biology, one idea is to study haplodiplontic cycles, in which, e.g., a gene manifested in a male will play a 1-player game, and the same gene in a female will play a 2-player game. The study of species with alternations of generations is another possible application of the ideas developed here. We also envisage an application in the study of parent-offspring conflict, in particular, the evolution of genomic imprinting, as the number of players changes during the development (for example comparing the number of players during gestational time and after that period). Another application is the study of the coexistence between one large individual of a given species with a cohort of individuals in a mutualistic or parasitic interaction.

\section*{Acknowledgement}
FACCC and MH thanks C. Gokhale for suggesting applying the ideas from~\citep{ChalubSouza_BMB19} to a non-fixed number of players. This idea is at the origin of the present work. MH and FACCC are funded by national funds through the FCT – Fundação para a Ciência e a Tecnologia, I.P., under the scope of the projects UIDB/00297/2020 (https://doi.org/10.54499/UIDB/00297/2020) and UIDP/00297/2020 (https://doi.org/10.54499/UIDP/00297/2020) (Center for Mathematics and Applications). FACCC also acknowledges the support of the project \emph{Mathematical Modelling of Multi-scale Control Systems: applications to human diseases}  2022.03091.PTDC (https://doi.org/10.54499/2022.03091.PTDC), supported by national funds (OE), through FCT/MCTES. (CoSysM3).
MH and FACCC contributed equally to the development of the work's ideas, computational codes, data analysis, discussions, and writing of the final version of the manuscript. We also thank both referees for many suggestions that helped to improve this work, and Renata Ramalho (Universidade NOVA de Lisboa) for helping in the language editing of the manuscript.


\begin{thebibliography}{36}
\providecommand{\natexlab}[1]{#1}
\providecommand{\url}[1]{\texttt{#1}}
\expandafter\ifx\csname urlstyle\endcsname\relax
  \providecommand{\doi}[1]{doi: #1}\else
  \providecommand{\doi}{doi: \begingroup \urlstyle{rm}\Url}\fi

\bibitem[Archetti(2009)]{Archetti_JTB}
M.~Archetti.
\newblock The volunteer's dilemma and the optimal size of a social group.
\newblock \emph{J. Theoret. Biol.}, 261\penalty0 (3):\penalty0 475--480, 2009.
\newblock ISSN 0022-5193.
\newblock URL \url{https://doi.org/10.1016/j.jtbi.2009.08.018}.

\bibitem[Axelrod(1984)]{Axelrod84}
R.~Axelrod.
\newblock \emph{The Evolution of Cooperation}.
\newblock Basic, New York, 1984.

\bibitem[Axelrod and Hamilton(1981)]{Axelrod_Hamilton}
R.~Axelrod and W.~D. Hamilton.
\newblock The evolution of cooperation.
\newblock \emph{Science}, 211\penalty0 (4489):\penalty0 1390--1396, 1981.
\newblock URL \url{https://doi.org/10.1126/science.7466396}.

\bibitem[Broom and Rychtář(2012)]{Broom_Rychtar2012}
M.~Broom and J.~Rychtář.
\newblock A general framework for analysing multiplayer games in networks using
  territorial interactions as a case study.
\newblock \emph{J. Theoret. Biol.}, 302:\penalty0 70--80, 2012.
\newblock ISSN 0022-5193.
\newblock URL \url{https://doi.org/10.1016/j.jtbi.2019.07.012}.

\bibitem[Carstensen et~al.(1999)Carstensen, Isaacowitz, and
  Charles]{Carstensen_AmPsy}
L.~Carstensen, D.~Isaacowitz, and S.~Charles.
\newblock Taking time seriously. a theory of socioemotional selectivity.
\newblock \emph{Am. Psychol.}, 54\penalty0 (3):\penalty0 165--81, 1999.
\newblock URL \url{https://doi.org/10.1037//0003-066x.54.3.165}.

\bibitem[Chalub and Souza(2014)]{ChalubSouza14a}
F.~A. C.~C. Chalub and M.~O. Souza.
\newblock {The frequency-dependent Wright-Fisher model: diffusive and
  non-diffusive approximations}.
\newblock \emph{J. Math. Biol.}, 68\penalty0 (5):\penalty0 1089--1133, 2014.
\newblock URL \url{https://doi.org/10.1007/s00285-013-0657-7}.

\bibitem[Chalub and Souza(2016)]{ChalubSouza_JMB16}
F.~A. C.~C. Chalub and M.~O. Souza.
\newblock Fixation in large populations: a continuous view of a discrete
  problem.
\newblock \emph{J. Math. Biol.}, 72\penalty0 (1-2):\penalty0 283--330, 2016.
\newblock ISSN 0303-6812,1432-1416.
\newblock URL \url{https://doi.org/10.1007/s00285-015-0889-9}.

\bibitem[Chalub and Souza(2017)]{ChalubSouza_JMB17}
F.~A. C.~C. Chalub and M.~O. Souza.
\newblock On the stochastic evolution of finite populations.
\newblock \emph{J. Math. Biol.}, 75\penalty0 (6-7):\penalty0 1735--1774, 2017.
\newblock ISSN 0303-6812,1432-1416.
\newblock URL \url{https://doi.org/10.1007/s00285-017-1135-4}.

\bibitem[Chalub and Souza(2018)]{ChalubSouza_JTB18}
F.~A. C.~C. Chalub and M.~O. Souza.
\newblock Fitness potentials and qualitative properties of the
  {W}right-{F}isher dynamics.
\newblock \emph{J. Theoret. Biol.}, 457:\penalty0 57--65, 2018.
\newblock ISSN 0022-5193,1095-8541.
\newblock URL \url{https://doi.org/10.1016/j.jtbi.2018.08.021}.

\bibitem[Chalub and Souza(2019)]{ChalubSouza_BMB19}
F.~A. C.~C. Chalub and M.~O. Souza.
\newblock From fixation probabilities to {$d$}-player games: an inverse problem
  in evolutionary dynamics.
\newblock \emph{Bull. Math. Biol.}, 81\penalty0 (11):\penalty0 4625--4642,
  2019.
\newblock ISSN 0092-8240,1522-9602.
\newblock URL \url{https://doi.org/10.1007/s11538-018-00566-w}.

\bibitem[Erovenko et~al.(2019)Erovenko, Bauer, Broom, Pattni, and
  Rychtář]{Erovenko_etal_2019}
I.~V. Erovenko, J.~Bauer, M.~Broom, K.~Pattni, and J.~Rychtář.
\newblock The effect of network topology on optimal exploration strategies and
  the evolution of cooperation in a mobile population.
\newblock \emph{P. Roy. Soc. A-Math. Phy.}, 475\penalty0 (2230):\penalty0
  20190399, 2019.
\newblock URL \url{https://doi.org/10.1098/rspa.2019.0399}.

\bibitem[Ferriere and Michod(2011)]{FerrierMichod_Nature2011}
R.~Ferriere and R.~E. Michod.
\newblock Inclusive fitness in evolution.
\newblock \emph{Nature}, 471\penalty0 (7339):\penalty0 E6--E8, 2011.
\newblock ISSN 0028-0836.
\newblock URL \url{https://doi.org/10.1038/nature09834}.

\bibitem[Fisher(1922)]{Fisher_1922}
R.~A. Fisher.
\newblock On the dominance ratio.
\newblock \emph{Proc. Royal Soc. Edinburgh}, 42:\penalty0 321--341, 1922.
\newblock URL \url{https://doi.org/10.1007/BF02459576}.

\bibitem[Gatchel(2021)]{Gatchel_2021}
M.~Gatchel.
\newblock Analyzing games with a variable number of players.
\newblock \emph{Proceedings of the AAAI Conference on Artificial Intelligence},
  35\penalty0 (18):\penalty0 15960--15961, May 2021.
\newblock URL \url{https://doi.org/10.1609/aaai.v35i18.17976}.

\bibitem[Gavrilets(2004)]{Gavrilets+2004}
S.~Gavrilets.
\newblock \emph{Fitness Landscapes and the Origin of Species (MPB-41)}.
\newblock Princeton University Press, Princeton, 2004.
\newblock ISBN 9780691187051.
\newblock URL \url{https://doi.org/10.1515/9780691187051}.

\bibitem[Gintis(2009)]{Gintis}
H.~Gintis.
\newblock \emph{Game theory evolving: a problem-centered introduction to
  modeling strategic interaction}.
\newblock Princeton University Press, Princeton, NJ, second edition, 2009.
\newblock ISBN 978-0-691-14051-3.

\bibitem[Gokhale and Traulsen(2010)]{GokhaleTraulsen_PNAS10}
C.~S. Gokhale and A.~Traulsen.
\newblock Evolutionary games in the multiverse.
\newblock \emph{P. Natl. Acad. Sci. USA}, 107\penalty0 (12):\penalty0
  5500--5504, 2010.
\newblock URL \url{https://doi.org/10.1073/pnas.0912214107}.

\bibitem[Herre and Wcislo(2011)]{HerreWcislo_Nature2011}
E.~A. Herre and W.~T. Wcislo.
\newblock In defence of inclusive fitness theory.
\newblock \emph{Nature}, 471\penalty0 (7339):\penalty0 E8--E9, 2011.
\newblock ISSN 0028-0836.
\newblock URL \url{https://doi.org/10.1038/nature09835}.

\bibitem[Hofbauer and Sigmund(1998)]{HofbauerSigmund}
J.~Hofbauer and K.~Sigmund.
\newblock \emph{Evolutionary Games and Population Dynamics}.
\newblock Cambridge University Press, Cambridge, UK, 1998.

\bibitem[Izquierdo et~al.(2014)Izquierdo, Izquierdo, and
  Vega-Redondo]{Izquierdo}
L.~R. Izquierdo, S.~S. Izquierdo, and F.~Vega-Redondo.
\newblock Leave and let leave: A sufficient condition to explain the
  evolutionary emergence of cooperation.
\newblock \emph{J. Econ. Dyn. Control}, 46:\penalty0 91--113, 2014.
\newblock ISSN 0165-1889.
\newblock URL \url{https://doi.org/10.1016/j.jedc.2014.06.007}.

\bibitem[Kurokawa(2019)]{Kurokawa_JTB19}
S.~Kurokawa.
\newblock Three-player repeated games with an opt-out option.
\newblock \emph{J. Theor. Biol.}, 480:\penalty0 13--22, 2019.
\newblock ISSN 0022-5193.
\newblock URL \url{https://doi.org/10.1016/j.jtbi.2019.07.012}.

\bibitem[Lessard(2011)]{Lessard_DGA11}
S.~Lessard.
\newblock On the robustness of the extension of the one-third law of evolution
  to the multi-player game.
\newblock \emph{Dyn. Games Appl.}, 1\penalty0 (3):\penalty0 408--418, 2011.
\newblock ISSN 2153-0785,2153-0793.
\newblock URL \url{https://doi.org/10.1007/s13235-011-0010-y}.

\bibitem[Maynard~Smith and Szathm{\'a}ry(1997)]{smith1997major}
J.~Maynard~Smith and E.~Szathm{\'a}ry.
\newblock \emph{The Major Transitions in Evolution}.
\newblock OUP Oxford, 1997.
\newblock ISBN 9780198502944.

\bibitem[Mayr(2001)]{ErnestMayr}
E.~Mayr.
\newblock \emph{What Evolution is}.
\newblock Basic Books, New York, 2001.

\bibitem[{Moran}(1962)]{Moran_1962}
P.~{Moran}.
\newblock \emph{{The statistical processes of evolutionary theory.}}
\newblock Clarendon, Oxford, 1962.

\bibitem[Nagylaki(1992)]{Nagylaki_1992}
T.~Nagylaki.
\newblock \emph{Introduction to theoretical population genetics}, volume~21 of
  \emph{Biomathematics}.
\newblock Springer-Verlag, Berlin, 1992.
\newblock ISBN 3-540-53344-3.
\newblock URL \url{https://doi.org/10.1007/978-3-642-76214-7}.

\bibitem[Nowak and Allen(2015)]{NowakAllen_PLOS2015}
M.~A. Nowak and B.~Allen.
\newblock Inclusive fitness theorizing invokes phenomena that are not relevant
  for the evolution of eusociality.
\newblock \emph{PLOS Biol.}, 13\penalty0 (4), 2015.
\newblock ISSN 1544-9173.
\newblock URL \url{https://doi.org/10.1371/journal.pbio.1002134}.

\bibitem[Nowak et~al.(2010)Nowak, Tarnita, and
  Wilson]{NowakTarnitaWilson_Nature2010}
M.~A. Nowak, C.~E. Tarnita, and E.~O. Wilson.
\newblock The evolution of eusociality.
\newblock \emph{Nature}, 466\penalty0 (7310):\penalty0 1057--1062, 2010.
\newblock ISSN 0028-0836.
\newblock URL \url{https://doi.org/10.1038/nature09205}.

\bibitem[Pacheco et~al.(2006)Pacheco, Traulsen, and Nowak]{Pacheco_2006}
J.~M. Pacheco, A.~Traulsen, and M.~A. Nowak.
\newblock Active linking in evolutionary games.
\newblock \emph{J. Theoret. Biol.}, 243\penalty0 (3):\penalty0 437--443, 2006.
\newblock ISSN 0022-5193.
\newblock URL \url{https://doi.org/10.1016/j.jtbi.2006.06.027}.

\bibitem[Skyrms and Pemantle(2000)]{Skyrms_Pemantle_PNAS2006}
B.~Skyrms and R.~Pemantle.
\newblock A dynamic model of social network formation.
\newblock \emph{P. Natl. Acad. Sci. USA}, 97\penalty0 (16):\penalty0
  9340--9346, 2000.
\newblock URL \url{https://doi.org/10.1073/pnas.97.16.9340}.

\bibitem[Souza et~al.(2009)Souza, Pacheco, and
  Santos]{Souza_Pacheco_Santos:JTB2009}
M.~O. Souza, J.~M. Pacheco, and F.~C. Santos.
\newblock Evolution of cooperation under $n$-person snowdrift games.
\newblock \emph{J. Theoret. Biol.}, 260\penalty0 (4):\penalty0 581--588, 2009.
\newblock ISSN 0022-5193.
\newblock URL \url{https://doi.org/10.1016/j.jtbi.2009.07.010}.

\bibitem[Taylor and Nowak(2006)]{TaylorNowak_2006}
C.~Taylor and M.~A. Nowak.
\newblock Evolutionary game dynamics with non-uniform interaction rates.
\newblock \emph{Theor. Pop. Biol.}, 69\penalty0 (3):\penalty0 243--252, 2006.
\newblock ISSN 0040-5809.
\newblock URL \url{https://doi.org/10.1016/j.tpb.2005.06.009}.
\newblock ESS Theory Now.

\bibitem[Taylor and Jonker(1978)]{TaylorJonker_1978}
P.~D. Taylor and L.~B. Jonker.
\newblock Evolutionarily stable strategies and game dynamics.
\newblock \emph{Math. Biosci.}, 40\penalty0 (1-2):\penalty0 145--156, 1978.
\newblock URL \url{https://doi.org/10.1016/0025-5564(78)90077-9}.

\bibitem[von Neumann and Morgenstern(2004)]{vonNeumann_Morgenstern}
J.~von Neumann and O.~Morgenstern.
\newblock \emph{Theory of games and economic behavior}.
\newblock Princeton University Press, Princeton, NJ, 2004.
\newblock ISBN 0-691-11993-7; 0-691-00362-9.
\newblock Reprint of the 1980 edition [Princeton Univ. Press, Princeton, NJ;
  MR0565457].

\bibitem[Wright(1931)]{Wright_1931}
S.~Wright.
\newblock {Evolution in Mendelian populations}.
\newblock \emph{Genetics}, 16\penalty0 (2):\penalty0 0097--0159, 1931.
\newblock ISSN 0016-6731.
\newblock URL \url{https://doi.org/10.1093/genetics/16.2.97}.

\bibitem[Zhaoyang et~al.(2018)Zhaoyang, Sliwinski, Martire, and
  JM]{Zhaoyang_PsyAging}
R.~Zhaoyang, M.~Sliwinski, L.~Martire, and S.~JM.
\newblock Age differences in adults' daily social interactions: An ecological
  momentary assessment study.
\newblock \emph{Psychol. Aging}, 33\penalty0 (4):\penalty0 607--618, 2018.
\newblock URL \url{https://doi.org/10.1037/pag0000242}.

\end{thebibliography}

\end{document}